\documentstyle[12pt]{article}
\newcommand{\be}{\begin{equation}}
\newcommand{\ee}{\end{equation}}
\begin{document}
\begin{titlepage}
\vspace{0.4cm}
\begin{center}
{\Large \baselineskip6pt {\bf On the violation of Marshall-Peierls
sign rule in the frustrated $J_{1}-J_{2}$ Heisenberg
antiferromagnet}\\}
\vspace{1cm}
\normalsize
{\it \baselineskip6pt J. Richter, N.B. Ivanov$^*$, and K. Retzlaff \\
\vspace{0.3cm}
\baselineskip6pt Institut f\"ur Theoretische Physik \\
Technische Universit\"at ''Otto von Guericke'' Magdeburg\\
Postfach 4120\\
D-39016 Magdeburg\\
Germany\\}
\end{center}

\vspace{0.5cm}

ABSTRACT

We present a number of arguments in favor of the suggestion that
the Marshall-Peierls sign rule survives the frustration in the
square-lattice Heisenberg antiferromagnet with frustrating
next-nearest-neighbor (diagonal) bonds ($J_{1}-J_{2}$ model) for
relatively large values of the parameter $J_{2}/J_{1}$. Both the
spin-wave analysis and the exact-diagonalization data concerning
the weight of Marshall states support the above suggestion.\\

\vspace{0.3cm}

\vspace{0.3cm}
\baselineskip6pt {$^*$ Permanent address: Georgi Nadjakov  Institute  for Solid
State Physics, Bulgarian Academy of Sciences, 72 Tzarigradsko chaussee blvd.,
1784 Sofia, Bulgaria}

\end{titlepage}

\newpage

    Many years ago Marshall proved, using a
Lemma due to Peierls, the well-known theorem determining the
amplitude phases of the set of Ising states building the ground-state
wave function of the spin $\frac{1}{2}$ Heisenberg antiferromagnet
on bipartite lattices [1]. Later on, Lieb and Mattis generalized
the theorem to arbitrary site spins and bipartite lattices without
translational symmetries [2,3].

Assigned to the $J_{1}-J_{2}$ model
\be\label{h}
H = H_{1}+\alpha H_{2} \equiv \sum_{nn} {{\bf S}_i{\bf S}_j} +
\alpha \sum_{nnn} {{\bf S}_{i}{\bf S}_j} \hspace{0.2cm},
\hspace{0.5cm} J_{1} \equiv 1 \hspace{0.5cm} \alpha \equiv
J_{2}/J_{1}=J_{2} \hspace{0.2cm},
\ee
($nn$ and $nnn$ mean that  the  summations run over  the
nearest-neighbor  and  next-nearest-neighbor, diagonal, bonds
on the square lattice, respectively), the theorem,
applicable to the case $\alpha \leq 0$, says:

(i) The lowest eigenstate of the Hamiltonian (1) in each subspace
determined by the eigenvalue $M$ of the spin operator $S_{total}^z$
reads
\be\label{gs}
\Psi_{M} = \sum_{m}{c_{m}^{(M)}|m \rangle} \hspace{0.2cm},
\hspace{0.5cm} c_{m}^{(M)}>0 \hspace{0.2cm}.
\ee
Here the Ising states $|m\rangle$ are defined by
\be\label{m}
|m\rangle \equiv (-1)^{S_A-M_A}|m_1\rangle \otimes |m_2 \rangle \otimes
\cdots \otimes |m_N\rangle \hspace{0.2cm},
\ee
where $|m_i\rangle,\hspace{0.2cm}i=1,\cdots,N$, are the eigenstates
of the site spin operator $S_{i}^{z}$ ($ -s_{i} \leq m_{i} \leq s_i$), $S_A=
\sum_{i\in A}s_i$, $M_{A(B)}=\sum_{i\in A(B)}m_i$, $M=M_A+M_B$.
The lattice consists of two equivalent sublattices $A$ and $B$.
$s_i\equiv s$, $i=1,\cdots,N$ , are the site spins.
The summations in Eq.(2) are restricted by the condition $\sum_{i=1}^N
m_i=M$, $N$ being the number of sites.

A direct result of the property $c_m^{(M)}>0$, valid for each $m$ from
the basis set (3), is the nondegeneracy of $\Psi_M$, since it is
impossible to build up other orthonormal states without using
negative amplitudes $c_m^{(M)}$.

(ii) The total spin of $\Psi_M$ is $S=|M|$.
Since each $M$ subspace contains only eigenstates with $S\geq |M|$,
a result of (ii) is the relation
\be\label{e}
E(S)<E(S+1) \hspace{0.2cm}, \hspace{0.2cm} S\geq 0 \hspace{0.5cm},
\ee
with $E(S)$ being the lowest energy eigenvalue belonging to $S$.
It is worth noticing that the property (ii) is characteristic for a
larger set of Hamiltonians (1), because the relation (i)$\Rightarrow$
(ii) can always be proved.

In what follows we address the model (1) with frustrating diagonal
bonds, i.e., $\alpha>0$. According to Lieb-Mattis definition [3], if
$\alpha>0$,
the square lattice is not bipartite  because it is not
possible to find a positive constant $c>0$ such that the conditions
$J_{AA}$, $J_{BB}\leq c$, $J_{AB}\geq c$ are satisfied
for {\bf all} bonds of the square
lattice. The goal of the present Letter is to give arguments in
favor of the suggestion that the Marshall-Peierls sign rule
(i) can survive the frustration for relatively large positive
parameters $\alpha$.

For references, let us firstly recall the important steps leading
to the sign rule (i) [3]. We suppose that the ground state of (1)
in the subspace $M$ (with an eigenvalue $E_M$)
 is writen in the form (2). The
Schr\"{o}dinger equation for the amplitudes $c_m^{(M)}$ reads
\be\label{sh1}
\sum_{n}A_{nm} c_n^{(M)}=(\varepsilon_m-E_M)c_m^{(M)} \hspace{0.5cm},
\ee
where
\be\label{me}
A_{nm}=B_{nm}-\alpha C_{nm} \equiv -\langle n|H_1^{xy}|m\rangle-
\alpha\langle n|H_2^{xy}|m\rangle \hspace{0.5cm},
\ee
\be\label{xxx}
B_{nm} \hspace{0.2cm}, C_{nm} \geq 0 \hspace{0.2cm}.
\ee
$\varepsilon_m$ ($\varepsilon_m > E_M$) is the eigenvalue of
$H^z$ in the Ising state $|m\rangle$.

Further, notice that the variational function $\Phi_M=\sum_m|c_m^{(M)}|
|m\rangle$, producing the Schr\"{o}dinger equation
\be\label{se2}
\sum_nA_{nm}|c_n^{(M)}|=(\varepsilon_m-E_M)|c_m^{(M)}| \hspace{0.2cm},
\ee
should also have an eigenvalue $E_M$. Let us suppose that the
matrix elements $A_{nm}$ satisfy
\be\label{nnn}
A_{nm}=B_{nm}-\alpha C_{nm} \geq 0.
\ee
Then Eqs.(5,8) lead to
\be\label{bb}
|\sum_nA_{nm}c_n^{(M)}|=\sum_n|A_{nm}||c_n^{(M)}| \hspace{0.2cm},
\ee
which is possible if, and only if, $c_m^{(M)} \geq 0$. Finally,
the chance of some $c_m^{(M)}$ to vanish is excluded, for if it were so,
Eqs.(5,10) would imply that $c_m^{(M)}=0$ for all $m$. These are
the main points leading to the sign rule (i).

Now, notice that for $\alpha>0$ the above reasoning is not
applicable since the inequality (9) will be violated for some
matrix elements $A_{nm}$. Nevertheless, for small enough
frustration parameters $\alpha >0$, one can argue that the
Marshall-Peierls sign rule survives the frustration. Indeed,
on rather general variational grounds the amplitudes $c_m^{(M)}$
are expected to be continuous functions of the parameter $\alpha$.
 The only point where one may expect some discontinuity of the amplitudes
 is the special point $\alpha =0$ where the system (1) acquires
a new quality, namely, frustrating bonds. So, let us assume that
 for some $m$ the amplitudes satisfy $c_m^{(M)}(\alpha=0^+)\leq 0$.
Then in the limit $\alpha \downarrow 0^+$, (9) will imply
\be\label{ccc}
\lim_{\alpha \downarrow 0^+} A_{nm} = B_{nm} \geq 0 \hspace{0.2cm},
\ee
showing that the inequality (9) is satisfied at $\alpha = 0^+$.
Thus, the same reasoning as in the case $\alpha \leq 0$ implies
that the above assumption is wrong, i.e., all the amplitudes
should be positive, $c_m^{(M)}(\alpha = 0^+)>0$.
Finally, since all the
possible Ising configurations take part in the ground state at
$\alpha = 0^+$ with finite positive amplitudes, it is clear
that the same property conserves in some region of positive
$\alpha$, as well.

{\it Spin-wave analysis}: The self-consistent spin-wave approach
 used below [4,5] is predominantly addressed to low-dimensional
spin systems. The conventional spin-wave technique is supplemented
with a condition for zero sublattice magnetization, thus
fulfilling, by hand, Mermin-Wagner's theorem [6] at finite
temperatures, or the same requirement on finite lattices.
The $J_1-J_2$ model has also been analyzed by this method
in a number of recent papers (for a review, see, e.g.,
Ref.7 and references therein).
Here we omit the details  and directly present the spin-wave
ansatz due to the theory mentioned above
\be\label{an}
\psi_{sw}  \hspace{4pt} \sim \hspace{4pt}
\exp {\Big(\! \sum_{\bf k}{\!
' w_{{\bf k}} \hspace{4pt}
\hat{a}^+_{{\bf k}} \hat{b}^+_{-{\bf k}}}  \Big)} \hspace{7pt}
|N\acute{e}el\rangle \hspace{0.5cm} .
\ee
Here $|N\acute{e}el\rangle$ is the classical N\'{e}el state. The  weight
 factors  $w_{\bf k}$ are defined by $w_{\bf k} = v_{\bf k} / u_{\bf k}$
    ,  $v_{\bf k}$ and  $u_{\bf k}$ being the  well-known
Bogoliubov coefficients
\be\label{bc}
  2v_{\bf k}^2 \, = \, {(1-\eta_{\bf k}^2 )}^{-1/2} \, - \, 1
  \hspace{1cm}, \hspace{1cm}  2u_{\bf  k}^2 \, =
  \, (1- \eta_{\bf k}^2  ) ^{-1/2} \,+ \, 1 \hspace{0.2cm},
\ee
\be\label{par}
\eta_{\bf k} = \frac{\gamma_{\bf k}}{ 1+ \mu - \alpha U (1-\Gamma_{\bf
k} )} \hspace{0.4cm}, \hspace{0.4cm}
 \gamma_{\bf k} = \frac{1}{2}( \cos{k_x} + \cos{k_y} )\hspace{0.4cm},
\hspace{0.4cm}
 \Gamma_{\bf k} = \cos{k_x} \cos{k_y} \hspace{0.3cm}.
\ee
The Bose operators $\hat{a}_{\bf R}$, $\hat{b}_{\bf R}$ come from
the Dyson-Maleev  transformation
and live on $A$ and $B$ sublattices,  respectively.  The  prime  over
sums means that $\bf k$ vectors run in the small Brillouin zone.
$\mu$ is the Lagrange multiplier used to imply the condition for
zero sublattice magnetization. The renormalization factor $U$,
renormalizing the frustation parameter $\alpha$ in the last
equation, is a result of the Hartree-Fock decoupling corresponding
to the theory (for details, see Refs.8-10). Within the theory,
$U$ is a ratio of two short-range correlators determined by the
 self-consistent equations. It has been shown in a recent study [10]
that the spin-wave ansatz (12) gives an exellent fit to the
exact-diagonalization
data for the relevant quantities of the model (1) up to
$\alpha \approx 0.45$ , provided the quasiclassical limit
(large site spin
$s_i \equiv s$) for the factor $U$ (see, e.g., Refs.11,12)
\be\label{u1}
U = \frac{1-0.102/2s+O[{(2s)}^{-2}]}{1+0.158/2s+O[{(2s)}^{-2}]}
\ee
is used. $U=1$ corresponds to the linear spin-wave approximation [13].

Now, let us rewrite (12) in the form
\be\label{psis}
\psi_{sw}  \hspace{4pt} \sim \hspace{4pt}
\exp {\Big(\! -  \sum_{{\bf R} , {\bf r} \atop {{\bf R}
 \in A}}
{w( {\bf r} ) \hat{a}_{\bf R}^{+} \hat{b}_{\bf R+r}^{+} }\Big)} \hspace{7pt}
|N\acute{e}el\rangle \hspace{0.2cm} ,
\ee
where the pairing function $w({\bf r})$ is defined by
\be\label{w}
w({\bf r}) \, = \, \frac{2}{N} \sum_{\bf k}{\! ' w_{\bf k} \cos{\bf kr}}
\hspace{0.2cm}.
\ee
The vector $\bf r$ connects sites from different sublattices.
{}From the structure of the latter state (16), it is clear that the
sign rule (i) breaks if, and only if, the pairing function
$w({\bf r})$ changes its sign for some vector $\bf r$
connecting two spins living on different sublattices.
 For the  $4 \times 4$  lattice this is just the vector
 ${\bf r} = \hat{\bf x}+2 \hat{\bf y}$ (and  the vectors related
 by the lattice symmetry ).
A numerical calculation demonstrates that
$w( \hat{\bf x}+2 \hat{\bf y})$
 changes  sign
at a point practically coinciding with the related $N = \infty$ case
($\alpha_M = 0.323$ for $U=1$). In the thermodynamic limit $N=\infty$,
we have checked numerically that it is just the function
$w( \hat{\bf x}+2 \hat{\bf y})$ which firstly becomes negative
when $\alpha$ increases. Unfortunately, we have not
succeeded in finding any analytical proof for this
observation. Finally, it is interesting to notice that the
curve representing the exact weight of Marshall states $vs$ $\alpha$
for the $4\times 4$ lattice is surprisingly well reproduced
from the finite-size spin-wave analysis based on the ansatz (16)
[10,14].

{\it Exact-diagonalization data}: Here we present
exact-diagonalization results for the weight of non-Marshall
states (which do not follow the sign rule (i)) in the exact
ground-state wave function of the $J_1-J_2$ model, Eq.(1),
on $4 \times 4$ and $6 \times 4$ lattices in the interval
$0 \leq \alpha \leq 0.52$, Fig.1 (A detailed and
varied analysis of that matter for the $S=0$ state is presented in Ref.14).
Here we study   the ground-state wave functions corresponding to
total-spin quantum numbers $S=0$, $1$, $2$, and $3$.
It is found a relatively wide
region where the non-Marshall states lack at all (for each
of the studied states). For example, the exact upper bound
$\alpha_M$ in the case $N=4\times 4$, $S=0$ is $\alpha_M= 0.28$
(this number seems to be slightly size-dependent as seen in
Fig.1, $6\times 4$ lattice).
It is worth noticing that even in the strongly frustrated
region $\alpha \approx 0.52$ the weight of the Ising states
violating the rule (i) is very small as compared to the
weight of states which fulfill the rule. Further, a well
pronounced tendency is that in the states with larger spin $S$
the weight of Marshall states is larger, too. In addition,
a more detailed
study also shows [14] that the exclusion of the non-Marshall
states from the exact ground states does not have any
drastic effect on the spin-spin correlators and the other
relevant quantities characterizing the model (1). It is
interesting to compare the above picture to the one
produced by the $J_1-J_2$ antiferromagnetic chain, Fig.2.
As it is clearly seen, the behavior is quite different: First,
the range where the sign rule is exactly fulfilled is
considerably reduced. Second, for states with larger spins
 the weight of non-Marshall states is larger, a tendency which is
just the opposite to the one seen in the two-dimensional case.
Precisely  at $\alpha = 0.5$ the exact ground state is known to be
the spin-Peierls dimer state, so that the
Marshall-Peierls sign rule is exactly fulfilled.

{\it Concluding remarks}:
In conclusion, a number of arguments has been given in favor
of the suggestion that in the $J_1-J_2$ model the Marshall-Peierls
sign rule survives the
frustration in a wide region up to the
limit $\alpha_M$. Evidently, at this point some
of the Ising amplitudes
in the ground state should vanish and further become negative.
This new quality of the ground state, however, does not
sufficiently influence the quantities characterizing the
long-range order in the system, which is rather natural in view
of the picture drawn above.
 It is worth noticing that
the number of  non-Marshall states just after the Marshall point
$\alpha_M$, seen in the exact results, is very small, which
is in agreement with the spin-wave picture showing that the
violation of the sign rule starts from Ising states with a
small number of reversed (in respect to the classical N\'{e}el state)
sublattice spins. Further, the weight of Ising states violating the
rule (i) remains extremly small up to the strong frustration limit
$\alpha \approx 0.5$. This fact has some importance for the construction of
variational wave functions, for which the rule (i) is applicable even in
the strongly frustrated region.

It is also
clearly seen that the ground states with larger spins $S$
are,  in a sense, less destroyed exactly in that case (two-dimensional
$J_1-J_2$ model) when
the ground state is, most probably, macroscopically  N\'{e}el ordered.
In addition, it is interesting to notice that the sharp increase in
the weight of non-Marshall Ising states with larger spins, Fig.1, is
observed close to the same point $\alpha \approx 0.5$ where,
according to the renormalized spin-wave ansatz (15,16) (see Ref.10),
the two-sublattice long-range Ne\'{e}l order is completely desrtoyed.
Therefore, there are some indications in favor of the suggestion that
the phenomenon discussed in the present Letter  may, in principal, be
 effectively used in  the finite-size
 analysis concerning the mechanism of breaking of
the global rotational symmetry of the Heisenberg Hamiltonians
 [15,16]. Clearly, further work in this direction is needed.
 For such studies, the
frustrated $J_1-J_2$ Heisenberg antiferromagnet seems to be quite
an adequate model, for the whole process of destroying of the N\'{e}el
order can
be suitably studied by changing the continuous  frustration parameter
$J_2/J_1$.
\begin{center}

                         * * *
\end{center}

This research
was  supported  by  Deutsche  Forschungsgemeinschaft,  Project  No
Ri 615/1-1, and  Bulgarian Science Foundation, Grant  $\Phi$2/91.
\newpage
{\Large {\bf References}} \\
1.Marshall W., Proc.Roy.Soc.A,{\bf 232}(1955)48.\\
2.Lieb E.H., Schultz T.D., and Mattis D.C., Ann.Phys.(N.Y.),
   {\bf 16}(1961)407.\\
3.Lieb E.H. and Mattis D.C., J.Math.Phys., {\bf 3}(1962)749.\\
4.Takahashi M., Phys.Rev.Lett., {\bf 58}(1987)168.\\
5.Takahashi M., Phys.Rev.B, {\bf 40}(1989)2494.\\
6.Mermin N.D. and Wagner H., Phys.Rev.Lett., {\bf 22}(1966)1133.\\
7.Chandra P., Coleman P., and Larkin A.I.,  J.Phys:
    Condens.Matter, {\bf 2}(1990)7933.\\
8.Xu J.H. and  Ting C.S., Phys.Rev.B, {\bf 42}(1990)6861.\\
9.Ivanov N.B. and Ivanov P. Ch., Phys.Rev.B, {\bf 46}(1992)8206.\\
10.Ivanov N.B. and Richter J., Phys.Rev.B, to be published.\\
11.Chubukov A., Phys.Rev.B, {\bf 44}(1991)392.\\
12.Bruder C. and Mila F., Europhys.Lett.,{\bf 17}(1992)463.\\
13.Hirsch J.E. and Tang S., Phys.Rev.B, {\bf 39}(1989)2887.\\
14.Retzlaff K., Richter J., and Ivanov N.B., Z.Phys., to be published.\\
15.Bernu B., Lhuillier C., and Pierre L., Phys.Rev.Lett.,
 {\bf 69}(1992).2590\\
16.Azaria P., Delamotte B., and Mouhanna D., Phys.Rev.Lett.,
 {\bf 70}(1993)2483.\\
\newpage
{\Large {\bf Captions of figures}} \\
{\bf Fig.1:} The weight of the non-Marshall Ising
states in the lowest-energy eigen
states with total spins $S=0,1,2,$ and $3$
$vs$ the frustration
parameter $J_2/J_1$ for the  $s=1/2$ square-lattice Heisenberg
antiferromagnet
with frustrating diagonal bonds for $4\times 4$ and $6 \times 4$ lattices.

{\bf Fig.2:} The weight of the non-Marshall Ising
states in the lowest-energy eigen
states with total spins $S=0,1,2,$ and $3$
$vs$ the frustration
parameter $J_2/J_1$ for the  $s=1/2$  Heisenberg
antiferromagnetic chain
with frustrating next-nearest neighbor bonds
for $N=16$ and $N=24$.
\end{document}